\documentclass[fleqn,usenatbib]{mnras}
\usepackage{newtxtext,newtxmath}
\usepackage[T1]{fontenc}
\DeclareRobustCommand{\VAN}[3]{#2}
\let\VANthebibliography\thebibliography
\def\thebibliography{\DeclareRobustCommand{\VAN}[3]{##3}\VANthebibliography}

\usepackage{graphicx}	% Including figure files
\usepackage{amsmath}	% Advanced maths commands
\usepackage{multirow}
\usepackage{xcolor}
\usepackage{amsmath}
\usepackage{array}
\usepackage{threeparttable}
\usepackage{subcaption} 
\title[Na-abundances of young thick disc stars]{Testing stellar yield prescriptions in OMEGA$+$: Implications for rising sodium abundances in young thick disc stars} %Current Galactic Chemical Evolution models fail to explain rising Na-abundances of young thick disc stars

\author[E. K. Owusu et al.]{Evans K. Owusu$^{1}{}^{,2}{}^{,3}{}^{,4}$,
Ashley J. Ruiter$^{1,2,3,4,5}$,
Alex J. Kemp$^{6}$,
Sven Buder$^{2}{}^{,7}$,
Ivo R. Seitenzahl$^{4,5,7}$,\newauthor
Nicolas Rodriguez-Segovia$^{1}$,
R. Pakmor$^{8}$,
Giulia C. Cinquegrana$^{2}{}^{,9}$,
Nicholas Storm$^{3}{}^{,10}$,\newauthor
Philipp Eitner$^{3}{}^{,10}$,
~and Maria Bergemann$^{3}$\\
$^1$School of Science, University of New South Wales Canberra, Australia Defence Force Academy, ACT 2600, Australia; \\
$^2$ARC Centre of Excellence for All-Sky Astrophysics in 3 Dimensions (ASTRO-3D)
\\
$^{3}$Max-Planck-Institut f\"{u}r Astronomie, K\"{o}nigstuhl 17, D-69117 Heidelberg, Germany;\\
$^{4}$Heidelberger Institut für Theoretische Studien, Schloss-Wolfsbrunnenweg 35, 69118 Heidelberg, Germany;\\
$^{5}$ Mathematical Sciences Institute, Australian National University, Canberra, ACT 0200, Australia;\\
$^{6}$Institute of Astronomy (IvS), KU Leuven, Celestijnenlaan 200D, 3001 Leuven, Belgium;\\
$^{7}$Research School of Astronomy and Astrophysics, Australian National University, ACT 2611, Australia;\\
$^{8}$Max Planck Institute for Astrophysics, Karl-Schwarzschild-Str.~1, D-85748 Garching, Germany;\\
$^{9}$School of Physics \& Astronomy, Monash University, Clayton, VIC 3800, Australia;\\
$^{10}$Heidelberg University, Grabengasse 1, 69117 Heidelberg, Germany;\\
}

\date{Accepted XXX. Received YYY; in original form 2025 June 21}
\pubyear{\the\year{}}

\begin{document}
\label{firstpage}
\pagerange{\pageref{firstpage}--\pageref{lastpage}}
\maketitle

% Abstract of the paper
\begin{abstract}
We recently identified an upturn in [Na/Fe] for the population of Solar-type stars in the Galactic young thick disc ($-0.3 < \mathrm{[Fe/H]} < +0.3$ dex) at super-Solar metallicity in GALactic Archaeology with HERMES (GALAH) data. Here, we investigate the cause of this unexplained Na enrichment ([Na/Fe] $\approx~0$ -- $0.6$ dex) using the OMEGA$+$ galactic chemical evolution code. We investigate the increase of [Na/Fe] with four combinations of nucleosynthetic yields from the literature, with source contributions from core-collapse supernovae, asymptotic giant branch stars, and Type Ia supernovae. We focus on two possible causes for the Na-enhancement: the `metallicity effect’ resulting from core-collapse supernovae at super-Solar metallicity and the contribution of metal-rich AGB stars. We adopt two sets of Type Ia supernova yields with one model assuming only Chandrasekhar-mass explosions, and another assuming only sub-Chandrasekhar-mass explosions. We find that the assumed Type Ia explosion has little effect on the [Na/Fe] Galactic Chemical Evolution modelling, and all Galactic chemical evolution models tested fail to reproduce the observed [Na/Fe] enrichment in the young thick disc population at super-Solar metallicities. Our study indicates a possible `under-pollution effect' by SNe Ia, which are the dominant producers of iron, in the Galactic disc's Solar-type star population. These findings provide a step forward toward understanding the origin of the unexplained sodium enrichment at super-Solar metallicities in the Galactic disc.
\end{abstract}


%%%%%%%%%%%%%%%%% BODY OF PAPER %%%%%%%%%%%%%%%%%%

\section{Introduction}
\label{sec:Intro} 
Stellar evolution is mainly determined by the initial chemical composition and mass of the star at birth \citep[see e.g.,][]{Appenzeller1982, Beccari2015, Hayden2017, Sellwood2002, Steinhardt2023}. In turn, the chemical composition of stellar nurseries is influenced by astrophysical processes such as stellar and galactic winds, supernovae (SNe), infalling gas and/or satellite galaxies, and star formation processes \citep{Gibson2003, Schoettler2024}. Consequently, Galactic Chemical Evolution (GCE) studies have been developed to trace the history and evolution of galaxies, as stellar abundances preserve the chemical signature of their birth environment and provide a crucial observational window into their evolution. 

All elements heavier than He and Li, from light elements like carbon to heavier ones like barium, are synthesised by one or more astrophysical channels: For example core--collapse supernovae (CC SNe), which synthesise large amounts of $\alpha$--elements such as Mg, Si, Ca, Ti; \citep{Arnett1978, Nomoto1997, Kobayashi2020, Kobayashi2020a}, Type Ia supernovae (SNe~Ia), the dominant source of iron--peak elements such as Cr, Mn, Fe, Co, Ni; e.g., \citep{Chiappini2009, Howell2011, Nomoto2013, Seitenzahl2013b, Leung2017, McWilliam2018, Bravo2024}, and asymptotic giant branch (AGB) stars, which produce most $s$--process elements, such as Ba, Sr, Y, and Zr; e.g, \citep{Karakas2007, Cseh2022, DenHartogh2023}. These three sites are some of the principal astrophysical sources driving GCE and are typically accounted for in GCE models. Additional (and more complex) sources include various types of interacting stars, in many cases undergoing one or more common envelope event(s), such as novae \citep{Kemp2024} and kilonovae \citep{Kobayashi2023, Matteucci2023}. 

Examining the contributions of different astrophysical phenomena toward the chemical enrichment of a galaxy over a wide range of metallicities provides insight into how Galactic processes shape enrichment of elements -- such as Na. The analysis of the enrichment pattern helps us trace how each channel influences the build-up of elements (e.g., Na) over time, ultimately shaping its present-day abundance relative to the Solar ratio. An analysis of the logarithmic abundance ratios of elements (X) relative to iron (Fe), expressed as [X/Fe]\footnote{\(\left[\text{X}/\text{Fe}\right] \equiv {\rm log}_{10}(N_\text{X}/N_\text{Fe}){_\star} - {\rm log}_{10}(N_\text{X}/N_\text{Fe}){_\odot} \)}, can provide valuable insights into the chemical evolution history of the Galaxy \citep{Francois1986, Wheeler1989, Freeman2002, Ratcliffe2022}. Iron is chosen as the reference element because its abundance increases steadily with time in the Universe, and it is relatively straightforward to determine its spectroscopic stellar abundances, particularly compared to elements whose gas-phase abundances are more challenging to measure directly. Furthermore, iron is produced and expelled back into the interstellar medium (ISM) by both CC SNe and SNe Ia (\citealt{Tsujimoto1995, Iwamoto1999, Kobayashi2006, RecioBlanco2014}), making it an excellent tracer of overall metallicity and GCE \citep{McWilliam1997, Jha2019}. This study focuses on Na, which is synthesised during hydrostatic carbon burning in massive stars (\(\mathrm{i.e.,}\,M_{\mathrm{ZAMS}}\ge8\,M_{\odot}\); \citealt{Woosley1995, Nomoto2013, Kobayashi2020a, Arcones2023}). Na is an element with one stable isotope (i.e., monoisotopic), and the next lightest element after Be and F. 

In massive stars, Na is synthesised during various nucleosynthetic processes. Some of this Na is expelled into the ISM before core collapse ensues, primarily via stellar winds. Alternatively, if Na is retained in the star, it can be ejected during the supernova explosion \citep{Vink2018}. Further, the amount of Na produced is sensitive to progenitor metallicity \citep[][i.e., there is a `metallicity effect']{Arnould1978}. Na's monoisotopic feature, that is, the isotope $^{23}\text{Na}$, which produces more neutrons than protons (11p,12n), makes its nucleosynthetic yields more sensitive to the neutron surplus existing in the progenitor star. Specifically, $^{23}\text{Na}$ production in massive stars during hydrostatic carbon and neon burning depends on the availability of free neutrons, which in turn is set by the initial metallicity of the star \citep{Arnett1971, Arnould1978, Arnett1996, Kobayashi2006}. At higher metallicity (Z), there is a greater abundance of $^{22}\text{Ne}$ -- produced from $^{14}\text{N}$ left over from the carbon-nitrogen-oxygen (CNO) cycle. The CNO chain requires these elements as catalysts, meaning that a higher initial Z results in a more efficient hydrogen burning, which then enhances the $^{14}$N to $^{22}$Ne reaction \citep{Woosley1995, Arnett1996}. $^{22}$Ne, being neutron-rich, acts as a reservoir and favours the production of $^{23}\text{Na}$ \citep{Couch1974}. Thus, the sensitivity of Na yields to metallicity arises from the specific nucleosynthetic pathway of $^{23}$Na and the metallicity-dependent production of $^{22}$Ne, rather than it being a general property of all monoisotopic elements. Comparatively, neighbouring even-Z elements like Mg do not posses this property because of their multiple stable isotopes -- their yields are less sensitive to the neutron excess (and hence progenitor metallicity) as production can shift from e.g. $^{24}\text{Mg}$ to $^{25}\text{Mg}$ or $^{26}\text{Mg}$ when metallicity changes \citep{Nomoto2013}. This dependence on progenitor metallicity allows Na to serve as an effective tracer of Galactic chemical evolution \citep{Venn2004}. In contrast, at lower metallicity, the initial abundances of CNO elements are reduced, leading to less $^{14}$N and consequently less $^{22}$Ne and neutron excess. This results in lower yields of $^{23}$Na \citep{Timmes1995, Kobayashi2020}.

Similarly, intermediate-mass--mass stars on the AGB play a significant role in Na synthesis, with their contribution strongly dependent on metallicity. At high metallicity, the second dredge-up efficiently mixes CNO -- processed material, particularly $^{14}$N, to the stellar surface early in the AGB phase. This process leads to increased secondary production of both nitrogen and Na as metallicity rises, and remains robust across a range of metallicities \citep{Ventura2013, Karakas2014, Cristallo2015}. In contrast, Na production is dominated by hot bottom burning (HBB) at lower metallicity. Here, the third dredge-up brings freshly synthesised carbon into the envelope, which is converted to Na via the Ne-Na cycle during HBB. Because HBB operates at higher temperatures in metal-poor stars, this pathway enables efficient primary ${^{23}}$Na production even when the initial metal content is low. Recent AGB models by \citet{Cinquegrana2022} confirm these trends: They show that ${^{23}}$Na yields increase with metallicity due to secondary production at high Z, while primary production via HBB remains significant at low Z (see their fig. 4). The usefulness of Na in the context of the `metallicity effect' has been discussed before; for a detailed discussion of the metallicity effect associated with Na abundance, see \citet{Arnould1978, Shi2004} and section 4.1 of \citep[][hereafter O24]{Owusu2024}.
%%%%%%%%%%%%%%%%%%%%%%%%%%%%%%%%%%%%%%%%%%%%%%%%%%%%

Historically, \citet{Cayrel1970} first reported evidence of Na enrichment in dwarf and giant stars. Subsequent studies (e.g., \citealt{Francois1986, Matteucci1986, Edvardsson1993, Mowlavi1999, Shi2004}) have confirmed a Na upturn and enrichment with increasing metallicity across various stellar populations (halo, bulge and disc). \citet{Cayrel1970} laid the essential groundwork for understanding Na production in massive stars. Later on, \citet{Francois1986} underscored the critical role AGB star progenitors play in Na enrichment, thereby expanding our understanding of Na's nucleosynthetic contribution across different stellar populations. Additionally, \citet{Takeda1994} found that A -- F supergiant stars exhibit an excess in Na with [Na/H] having values ranging from $0.0 \leq \text{[Na/H]} \leq 0.5$ in F supergiants and $0.7 \leq \text{[Na/H]} \leq 0.8$ in A supergiants. This enhancement is credited to the Ne-Na cycle during hydrogen burning. More recent high-resolution spectroscopic surveys have demonstrated that this trend is particularly pronounced at super-Solar metallicities ([Fe/H] $> 0$), where [Na/Fe] shows a clear increase in both old and young Milky Way disc stars (see $024$). The Na enrichment at super Solar [Fe/H] is especially notable in the metal-rich, dynamically cold younger-thick disc population, a trend that has been reported in other studies \citep{Bensby2014, Bensby2017, Nissen2020, Griffith2022}. In particular, \citet{Bensby2014} documented an increase in [Na/Fe] with [Fe/H] $> 0$ for a sample of $714$ dwarf stars in the Solar neighbourhood (see their fig. $16$). 
Despite these reports of an upturn in Na enrichment at high metallicity, a universally accepted explanation remains elusive. These differing interpretations of the observed increase in Na, as well as contributions from different nucleosynthesis channels, motivated us to investigate the progressive rise in [Na/Fe] at high metallicity. It presents important implications for understanding nucleosynthesis sources and GCE. Furthermore, Na enrichment is reported in other stellar populations, particularly in globular clusters. In these environments, Na enrichment serves as a pivotal diagnostic for delineating their complex nucleosynthetic histories. A prominent feature is the Na-O anticorrelation, first reported by \citet{Cohen1978}, which has since been extensively investigated in subsequent work \citep{Sneden1997, Lee2010}. This anticorrelation manifests as a bipolar clustering of first and second generation stars, where second generation stars are identifiable by their enhanced Na and depleted O abundances. \citep{Carretta2009, Charbonnel2016, Carretta2019, Piatti2020}. This chemical signature is widely interpreted as the result of proton-capture reactions occurring at high temperature ($\approx$ T $\geq 4\times 10^7$ K) within the interiors of polluter stars from the previous generation, such as high-mass AGB stars and massive stars which subsequently eject processed material into their intracluster environment \citep{Decressin2007, Ventura2009}. The presence of Na anomalies, therefore, not only illuminates self-enrichment mechanisms operating with globular clusters but also provides vital insights into their enrichment processes. In addition, these anomalies offer crucial constraints on theoretical models of GCE. Together, these processes shape the build-up of light element abundances throughout the Galaxy over cosmic time \citep{Kobayashi2011}.

This paper is organised into five sections. In Section $2$, we describe the data and the processing of yield tables for the various theoretical stellar yields employed in this study. Additionally, we outline the choice of GCE code used. Section $3$ presents the results of this work, while Section $4$ discusses these results and their implications for broader nucleosynthesis processes. Finally, Section $5$ concludes with a summary of our findings and highlights opportunities for future research.

\section{Chemical Evolution Data and Modelling Methodology}\label{sec:data}
%%%%%%%%%%%%%%%%%%%%%%%%% Figure 1
\begin{figure*}
	\includegraphics[width=\textwidth]{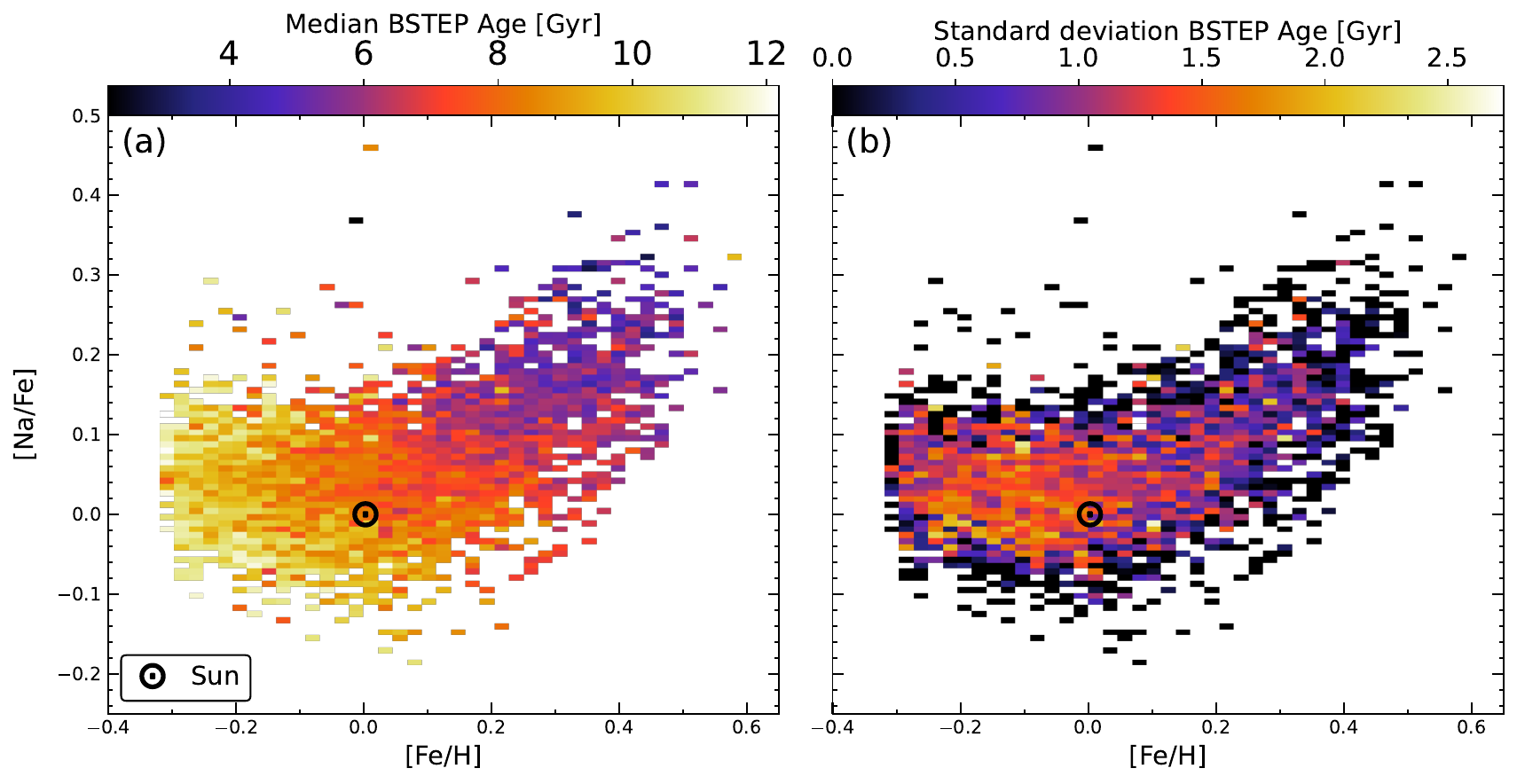}
     \caption{[Na/Fe] as a function of metallicity [Fe/H] for the sample of Solar-type stars from GALAH DR3 used in this work (see Section \ref{sec:data}). The left panel (a) shows the distribution coloured by stellar age, illustrating the median stellar age in (Gyr) at each ([Fe/H], [Na/Fe]) bin. The right panel displays the same distribution, highlighting the standard deviation in stellar age within the corresponding bins. The Bayesian Stellar Parameters Estimator (BSTEP) was used to compute the stellar ages. The Solar abundance position is marked by the symbol $(\odot)$ at [Fe/H], [Na/Fe] $= (0,0)$.}
    \label{fig:figure1}
\end{figure*}
%%%%%%%%%%%%%%%%%%%%%%%%%
In the third data release of the GALactic Archaeology with HERMES spectrograph \citep[GALAH DR3,][]{Buder2021}, Na is one of the $30$ elemental abundances analysed. The dataset includes Na lines at $\lambda4751.82\text{\AA}$, $\lambda5682.63\text{\AA}$ and $\lambda5688.21\text{\AA}$, with abundances derived using one--dimensional, non--local thermodynamic equilibrium (1D NLTE) modelling \citep{Amarsi2020}. In Figure \ref{fig:figure1}, we show the GALAH DR3 [Na/Fe] abundances for a sample of $5\,148$ Solar-type stars (young thick disc only, see section \ref{sec:sample} for definition), similar to what was discussed in O24, but with a higher metallicity range in the sample up to $+0.6$ dex to incorporate an extended population in metallicity-space. The reason for the observed [Na/Fe] upturn in the GALAH data is currently unknown and is the focus of this paper. This work aims to investigate the underlying cause of Na abundance enrichment at super-Solar metallicities (up to $0.6$ dex) among the younger population of the old (thick) disc stars of the Galaxy. 

\begin{figure}
	\includegraphics[width=\columnwidth]{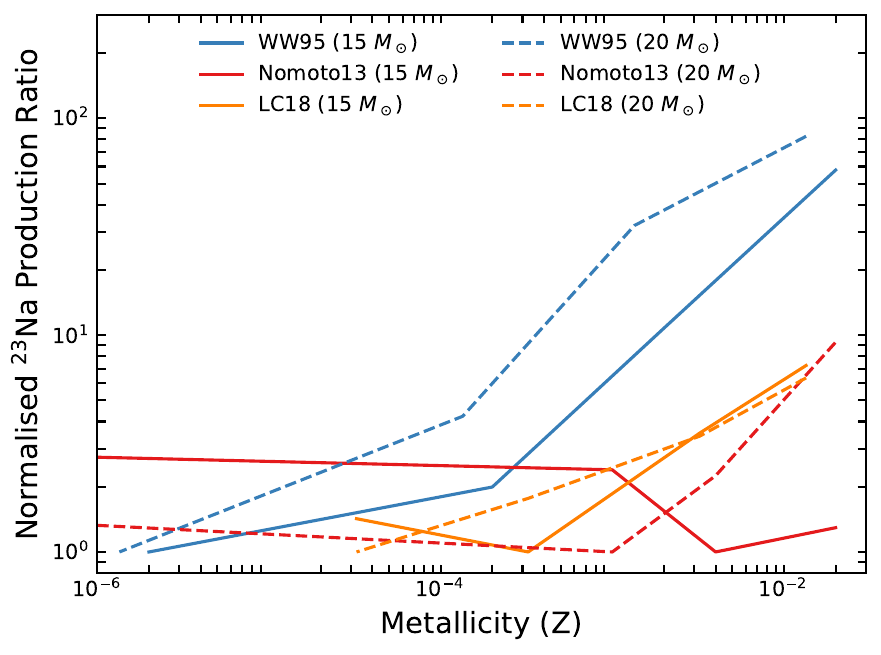}
     \caption{Normalised sodium ($^{23}$Na) production ratios as a function of stellar metallicity (Z) for massive stars with initial masses $15$ and $20$ M$_{\odot}$. Results are shown for models from \citet[WW95, blue]{Woosley1995}, \citet[Nomoto13, red]{Nomoto2013} and \citet[LC18, orange]{Limongi2018}. For each model, the $^{23}$Na yield at each metallicity is divided by the minimum yield across all metallicities, highlighting the relative change in Na production with metallicity.}
 \label{fig:figure2}
\end{figure}

We create four GCE models (see Table \ref{tab:model}) that incorporate different stellar yield predictions from theoretical nucleosynthesis calculations of CC SNe, SNe~Ia, and AGB stars and compare these with observational data.  
We explore two possibilities for the Na enhancement: the `metallicity effect' from CC SNe at high metallicity and AGB stars at super-Solar metallicity, as possible solutions for explaining the observed Na enhancement. Firstly, the metallicity effect from CC SNe motivated us to examine normalised Na production ratios using data from \citet{Woosley1995}. A preliminary check into those models (see Figure \ref{fig:figure2}) seemed to support the idea that the metallicity effect plays an important role in Na production. We therefore expand our yield table sets (see Table \ref{tab:model}) to include massive star yields from \citet{Woosley1995} in addition to existing massive star yield tables in OMEGA$+$ (i.e., \citealt{Kobayashi2006, Limongi2018}). Secondly,  we include \citet{Cinquegrana2022} mass- and metallicity-dependent AGB yields, which cover a broad range of super-solar metallicities (\(0.04 \leq \text{Z} \leq 0.1\)), into our chemical evolution models (see Section \ref{sec:stellar_yields} and Table \ref{tab:model} for model K10C22K06). The super-Solar metallicity AGB models clearly show an increasing $^{23}$Na yield with increasing metallicity (see their fig. 8). By incorporating both of these additional yield sets, we account for Na enrichment from both CC SNe and AGB channels.
%%%%%%%%%%%%%%%%%%%%%%%%%%%%%%%%% Figure 2
%%%%%%%%%%%%%%%%%%%%%%%%%%%%%%%%%%%%%%%
\subsection{Sample selection}
\label{sec:sample}

Similar to O24, we selected a sample of Solar-type (G spectral type) i.e., youngest stars within the thick disc with age $\sim 6 - 8$ Gyr (hereafter ``young thick disc'') from the GALAH DR3 catalogue with quality cuts based on the following criteria:
\begin{align} 
    \texttt{flag\_sp} == 0,
    \texttt{flag\_fe\_h} == 0, \texttt{flag\_Na\_fe} == 0, \label{eq:flags} \\
    \texttt{e\_age\_bstep}/\texttt{age\_bstep} < 0.2 \text{ or }\texttt{e\_age\_bstep} < 2, \label{eq:age_cuts} \\
    \texttt{snr\_c2\_iraf} > 50, \label{eq:snr_ccd2}\\
    5600 < T_\text{eff} < 5950, \log g > 4.15, \label{eq:effect_tem}\\
    -0.3 < \mathrm{[Fe/H]} < +0.6. 
    \label{eq:Z_range} 
 \end{align} 
\label{eq:FeH_range}
Equation \ref{eq:flags} describes the basic quality criteria applied to our sample. Equation \ref{eq:age_cuts} selects stars with a reliable age estimate, specifically those with uncertainties below $20\%$ or less than $2$ Gyr. Equation \ref{eq:snr_ccd2} ensures high-quality spectra by requiring a signal-to-noise ratio greater than $50$ per pixel in the green CCD \citep{Kos2017}. Finally, in Equation \ref{eq:effect_tem} and \ref{eq:Z_range}, we place additional constraints to select only Solar-type stars. In addition, we also apply the standard O24 selection criterion for young thick disc stars only. \(\text{Old stars}: [\text{Na}/\text{Fe}]_{\text{old}} > 0.55 - (0.5/5.5)\times\tau\); where $\tau$ is the stellar age in Gyr. However, compared to O24, we extended the metallicity range beyond [Fe/H] = $+0.3$ to capture a broader abundance distribution (see Equation \ref{eq:Z_range}). This wider range enables us to gain insights into Na enhancement trends. Following these criteria, our final sample consists of $5\,148$ Solar-type young thick disc stars, as illustrated in Figure \ref{fig:figure1}, which displays the resultant sample on a [Fe/H] vs [Na/Fe] plane. Significantly, we observe a persistent rising trend commencing at [Fe/H] $>0$.

\subsubsection{Spatial Coverage}

\begin{figure*}
	\includegraphics[width=\textwidth]{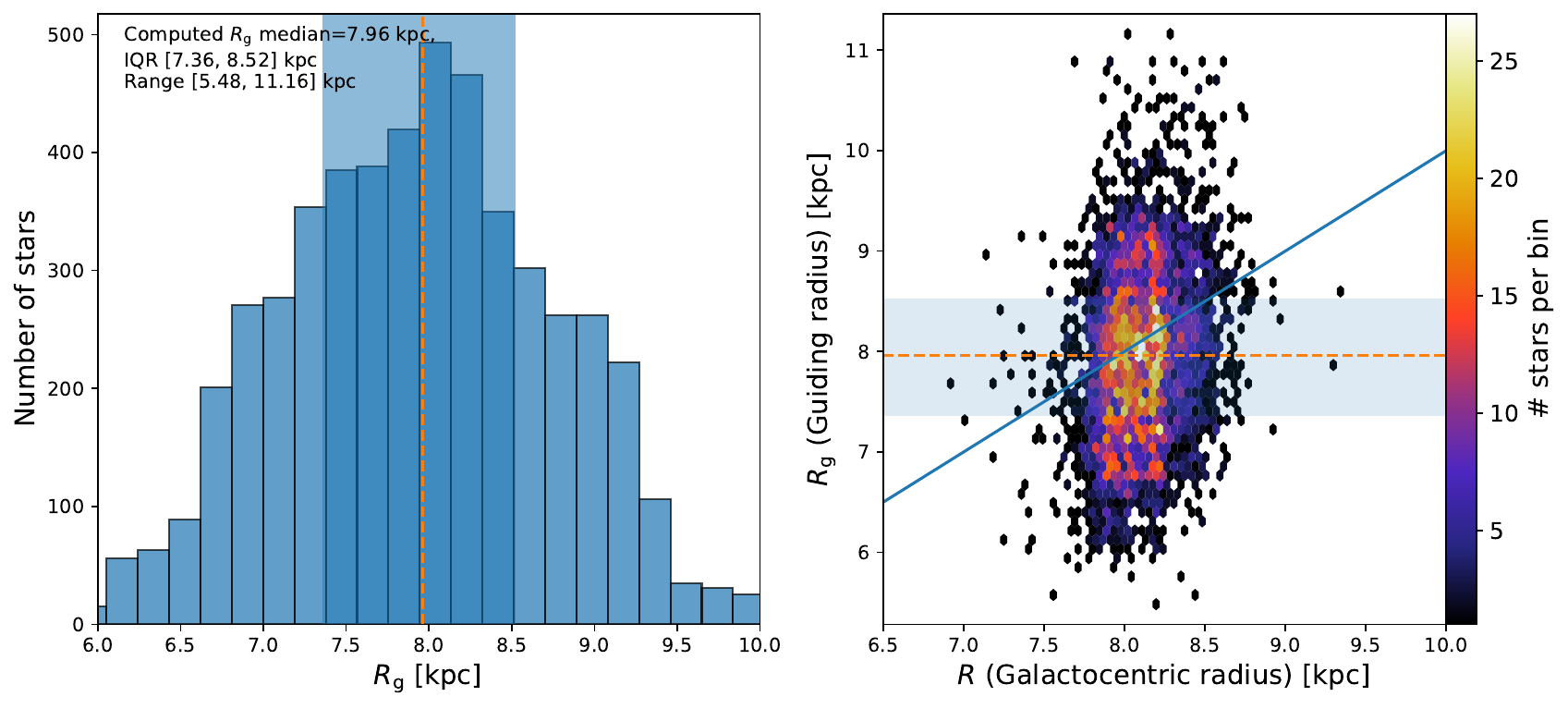}
     \caption{Spatial coverage of the young thick disc sample in the Solar neighbourhood. \textit{Left:} Histogram of guiding radii, $\mathrm{R_g}$, for the Solar-type star sample. The vertical dashed line marks the median ($\mathrm{R_g} = 7.96$ kpc); the blue band shows the interquartile range (IQR; $7.36 - 8.52$ kpc). The full extent of the distribution is $5.48 - 11.16$ kpc. \textit{Right:} Density map of guiding radius versus current Galactocentric radius ($\mathrm{R_g}$ vs. R). The solid diagonal line indicates the one-to-one relation ($\mathrm{R_g} =~ $R). The horizontal dashed line and shaded band represent the median and IQR from the left panel. The colour bar indicates the number of stars per bin.}
 \label{fig:Figure6}
\end{figure*}

While our observational data only represent a sub-component of the Galaxy's young thick disc and not the entire Galaxy, our sample covers a well-defined spatial range centred on the Solar neighbourhood, with a current Galactocentric radius and guiding radius median of $7.96$ kpc, an interquartile range of $7.36 - 8.52$ kpc, and an overall range extending from $5.48$ to $11.6$ kpc (see Figure \ref{fig:Figure6}). This coverage overlaps substantially with the ``canonical'' thick-disc selection of \citet{Duong2018}, who focused on the range of $7.9\leq\mathrm{R_g}\leq 9.5$ kpc to avoid contamination by inner-disc populations deliberately. Consequently, the chemical dispersion in our sample is dominated by local evolutionary history rather than the broader radial chemical gradient observed when stars with $\mathrm{R_g} < 6$ kpc are included \citep{Akbaba2024}. Our coverage thus predominantly samples the young thick disc population in this spatial window, from the inner disc to the onset of the outer disc. The goal of the study is to determine whether the observed Na enhancement at super-Solar metallicity can be, to first order, reproduced with a GCE code. While we assume that a single evolutionary/enrichment sequence characterises our sample, we acknowledge that radial migration plays a role in bringing metal-rich stars from the inner disc to the Solar neighbourhood. As such, the most metal-rich Na-enhanced stars may represent a distinct evolutionary path, shaped by radial migration and chemical enrichment histories different from the local young thick disc. This interpretation is consistent with recent observational studies that find old, metal-rich stars in the Solar vicinity likely migrated from smaller Galactocentric radii, thus reflecting a complex chemical evolution influenced by radial mixing \citep[e.g.,][]{Schonrich2009, Ness2016, Feuillet2018, Hayden2022}. 
%%%%%%%%%%%%%%%%%%%%%%%%%%%%%%%%%%%%%%%%%%%%%%%%%%%%%%%%%%%%%%%%%%%%%%%%%%%%%%
%%%%%%%%%%%%%%%%%%%%%%%%%%%%%%%%%%%%%%%%%%%%%%%%%%%%%%%%%%%%%%%%%%%%%%%%%%%%%% 
\subsubsection{BSTEP Stellar Ages}
We adopt ages from the GALAH Bayesian Stellar Parameters estimator (BSTEP) code, a Bayesian isochrone-fitting framework in which spectroscopic T$_\mathrm{eff}$, $\log$$g$, and [Fe/H] together with photometry and parallaxes are compared to stellar evolution grids to infer posterior probability distributions (PDFs) for age, mass and distance. The method follows the Bayesian formulation of \citet{Pont2004} and the GALAH implementation described by \citet{Sharma2018} and \cite{Buder2021}. We used the posterior medians as point estimates and the 16th-84th percentiles as credible intervals. As discussed in those works, inferred ages can shift with the adopted temperature and metallicity scales, isochrones, and priors; our conclusions do not rely on the absolute age scale and are robust to first-order systematics.
%%%%%%%%%%%%%%%%%%%%%%%%%%%%%%%%%%%%%%%%%%%%%% used the two-zone GCE code
\subsection{Chemical evolution with OMEGA$+$}
We model the chemical enrichment of the Galaxy using OMEGA$+$\footnote{Publicly available at https://github.com/becot85/JINAPyCEE} \citep{Cote2018}, an extension of the One-zone Model for the Evolution of Galaxies \citep[OMEGA,][]{Cote2017, Cote2018a}. Unlike the one-zone version, which treats the entire system as a single homogeneous reservoir, OMEGA$+$ employs a two-zone formalism that explicitly distinguishes between the primary star-forming regions (the Galaxy) and a surrounding circumgalactic medium (CGM). This structure allows a more flexible exploration of feedback-regulated star formation and the fate of metals removed from the ISM.

The list of some parameters in OMEGA$+$ capabilities that are relevant for this study, relative to one-zone models, are:
\begin{itemize}
\item[$\bullet$]Self-consistent star formation rate (SFRs): The star formation history is not imposed externally, but instead emerges from an equilibrium between inflows, gas consumption, and stellar feedback. The instantaneous SFR follows the Kennicutt-Schmidt law, \citep{Kennicutt1998}. 
\item[$\bullet$] Metal recycling via the CGM reservoir: Outflows do not irreversibly expel metals from the system. Instead, metals ejected by CC SNe, SN Ia and AGB winds are deposited into the CGM reservoir. These can later return to the Galaxy as recycled inflows on a timescale governed by the dynamics of gas exchange \citep{Cote2018a}. This treatment contrasts with the one-zone model, where metals are either locked into stars or permanently lost.

\item[$\bullet$] Flexibility in inflow composition: While inflow is often assumed to be of {\em primordial composition} for comparison with classical GCE studies \citep[e.g.,][]{Kobayashi2006, Matteucci2012}, the OMEGA$+$ framework permits inflows of enriched gas. In particular, a subset of the CGM can be preferentially enriched by CC SNe ejecta, as suggested by both theoretical models of galactic winds \citep{MacLow1999, Tumlinson2017, Fielding2017} and observations of galactic halos showing metal loading \citep{Werk2014, pereira2019search}. The metallicity of the re-accreted inflow is therefore an adjustable parameter.
\end{itemize}

OMEGA$+$ shares similarities with other GCE codes, such as the Versatile Integrator for Chemical Evolution \citep[VICE,][]{Johnson2021}, the flexible one-zone chemical evolution code \citep[FlexCE,][]{Andrews2017}, the Bayesian chemical evolution code ChemPy \citep{Rybizki2017}, and the Galactic Evolution tool \citep[GETool,][]{Fenner2002}.
In OMEGA$+$, gas flows are treated with two key assumptions. First, we assume that inflows are predominantly primordial in composition. Second, we model outflows as metal-enriched, allowing their metallicity to vary over time. These assumptions directly impact the predicted chemical evolution of the stimulated galaxy, shaping the enrichment history and the metallicity distribution of stars. OMEGA$+$ can better reproduce observed features such as the metallicity distribution function (MDF) and abundance patterns in stellar populations \citep{cote2016} by allowing the metal loading of outflows to change with time. In contrast, assuming constant metal loading or ignoring enriched outflows would result in significant discrepancies between the model predictions and observed chemical signatures. Table \ref{tab:stellar_params} outlines our adjusted physical parameters. The GCE model was evolved for $13$ Gyr. 

 \begin{table}
    \centering
    \caption{Yield tables used and the resulting models. The `model' reflects the combined selections of AGB and massive stars in each nucleosynthesis yield table (see Section \ref{sec:stellar_yields} for details of each model). Chandrasekhar-mass ($\text{M}_{\text{Ch}}$) Type Ia supernovae yields are adopted from S13 and sub-Chandrasekhar-mass (sub--$\text{M}_{\text{Ch}}$) Type Ia supernova yields are adopted from P22. Models are explained in the footnote.}
    % \begin{tabular*}{\columnwidth}{@{}l@{\hspace*{50pt}}l@{\hspace*{50pt}}l@{}{\hspace*{50pt}}l@{}}
%     \hline
%     AGB                                 & Massive star                          & SN\,Ia & Model      \\
%     \hline
%     Karakas 2010 (K10)                  & Limongi \& Chieffi 2018 (LC18)        & S13/P22 & K10LC18    \\[2pt]
%     K10                                 & Woosley \& Weaver 1995 (WW95)         & S13/P22 & K10WW95    \\[2pt]
%     K10                                 & Kobayashi 2006 (K06)                  & S13/P22 & K10K06     \\[2pt]
%     K10, Cinquegrana 2022 (C22)         & Kobayashi 2006 (K06)                  & S13/P22 & K10C22K06  \\[2pt]
%     \hline
%   \end{tabular*}
\resizebox{\columnwidth}{!}{%
  \begin{tabular}{llll}
  \hline
  AGB                                 & Massive star                          & SN\,Ia  & Model       \\
  \hline
  K10                  & LC18       & S13/P22 & K10LC18     \\
  K10                                 & WW95        & S13/P22 & K10WW95     \\
  K10                                 & K06                  & S13/P22 & K10K06      \\
  K10, C22         &K06                  & S13/P22 & K10C22K06   \\
  \hline
  \end{tabular}%
}

\begin{flushleft}
{\footnotesize$^1$ LC18: \citet{Limongi2018}; WW95: \citet{Woosley1995}; K06: \citet{Kobayashi2006}; K10: \citet{Karakas2010}; C22: \citet{Cinquegrana2022}; S13: \citet{Seitenzahl2013} and P22: \citet{Pakmor2022}.} 
\end{flushleft}
 \label{tab:model}
\end{table}

%%%%%%%%%%%%%%%%%%%%%%%%%%%%
Type Ia supernovae (SNe Ia) are crucial in GCE models as they dominate the production of iron-peak elements (e.g., Fe, Mn, Ni), contributing $\sim 70\%$ of the Milky Way's iron inventory \citep{Matteucci2009, Fink2012, Bovy2015, Maoz2018, Jha2019}. Unlike CC SNe, which primarily enrich the ISM with $\alpha$--elements such as O and Mg on short timescales ($\sim$Myr), SNe Ia inject iron with a delay time spanning $\sim 50$~Myr to a Hubble time. This delay creates the observed downturn in [$\alpha$/Fe] ratios as [Fe/H]$~>~-1$ in stellar populations, distinguishing chemically distinct Galactic components (e.g., thick vs. thin disc, \citealt{Matteucci2009, Dubay2024, Cavichia2024}). SNe Ia also produce more iron per event than CC SNe, making them indispensable for reproducing the G--dwarf metallicity distribution and explaining Solar abundance patterns \citep{Kobayashi2020}. To first order, SNe~Ia can be split into two categories when considering their nuclear yields: Chandrasekhar- and sub-Chandrasekhar-mass explosions. Given that both categories of SNe produce distinct elemental yields in varying amounts and often with different injection timescales, and given the ongoing uncertainty surrounding the nature of SN~Ia progenitors, we incorporate yields from one of each broad category into our GCE models. Though it is widely agreed that various formation channels of Type Ia supernovae are predicted to have different delay time distributions \citep[DTD, see][]{Ruiter2011, Eitner2023}, as a first step, we adopt the DTD from \citet{Maoz2012}, which is the default implementation in OMEGA$+$.
%%%%%%%%%%%%%%%%%%%%%%%%%%%%%%%%%%% Table 2
\begin{table}
\begin{center}
    \caption{Default and adjusted stellar parameters used in this work for the OMEGA$+$ GCE model. The star formation efficiency is a dimensionless scaling quantity and a free parameter (see Equations $13$ and $14$ of \citealt{Cote2018})} %.
% \begin{tabular}{ccc}
% \hline \hline
% Stellar parameters & Default values & New parameter \\ 
% \hline
% Initial mass function (IMF)  &1 -- 50 M$_{\odot}$& 1 -- 50 M$_{\odot}$ \\ 
% Star formation efficiency (SFE, $f_{\star}$) & $3.0 \times 10^{-9}$~yr$^{-1}$ & $3.0\times10^{-10}$~yr$^{-1}$ \\ 
% Initial mass of gas in simulation (mgal)         & 1.6e11 M$_{\odot}$                & $2.0 \times 10^{11}$ M$_{\odot}$ \\ 
% \hline \hline
% \end{tabular} $\times 10^{11}$ 

% \begin{tabular}{lcccrr}
% \hline \hline
% Stellar parameters & Default values & Adjusted values \\
% \hline
% Initial mass function (IMF)  & 1 -- 30 M$_{\odot}$ & 1 -- 130 M$_{\odot}$ \\[2pt]
% \hline
% Star formation efficiency (SFE; $\epsilon_{\star}$) & $3.0 \times 10^{-9}$ & $3.0 \times 10^{-10}$ \\[2pt]
%  \hline
% Initial mass of gas in simulation (mgal) & $1.0\times10^{11}$ M$_{\odot}$ & $2.0 \times 10^{11}$ M$_{\odot}$ \\[2pt]
% \hline \hline
% \end{tabular}
\resizebox{\columnwidth}{!}{%
  \begin{tabular}{llll}
  \hline
  Stellar parameters & Default values                & Adjusted values               \\
  \hline
  Initial mass function (IMF)  
    & 1 -- 30\,M$_\odot$              & 1 -- 130\,M$_\odot$            \\[2pt]
  
  Star formation efficiency (SFE; $\epsilon_{\star}$)  
    & $3.0\times10^{-9}$             & $3.0\times10^{-10}$           \\[2pt]
  
  Initial mass of gas in simulation (mgal)  
    & $1.0\times10^{10}$\,M$_\odot$   & $2.0\times10^{11}$\,M$_\odot$  \\[2pt]
  \hline
\end{tabular}%
}
\label{tab:stellar_params}
\end{center}
\end{table}

For simplicity, we assume {\em either} Chandrasekhar--mass delayed detonations or sub-Chandrasekhar-mass prompt detonations in white dwarf mergers, and apply the \citet{Maoz2012} DTD for both cases. For the Chandrasekhar mass explosions, we use the N$100$ model yields from \citet{Seitenzahl2013} for 4 metallicities (Z $= 0.0002$, $0.002, 0.01$ and $0.02$). For the sub-Chandrasekhar mass explosions, we use yields from the two-explosion double white dwarf merger model of \citet{Pakmor2022}, currently one of the most favoured scenarios for SNe~Ia \citep[see][for a recent review]{Ruiter2025}. Since only Solar metallicity yields are available for the merger model, we adopt these for the entire metallicity range sampled in our GCE study when the sub-Chandrasekhar mass scenario is assumed. 
%%%%%%%%%%%%%%%%%%%%%%%%%%%%%%%%%%%%%%%%%%%%%%%%%%

\subsubsection{Parameters for the GCE code}
Table \ref{tab:stellar_params} presents both default and adjusted stellar parameters used in OMEGA$+$ for this study. The default settings of OMEGA$+$ produce [Fe/H] values only up to approximately $0.2$ dex. To model higher metallicities, we modified these default parameters (see Section~\ref{sec:MWMC}), allowing the machinery to reach [Fe/H] values at higher super-Solar metallicities (i.e., [Fe/H] ${\sim}0.6$ dex), which overlaps the AGB star yields in one of our models (see K10C22K06 in Table \ref{tab:model}), enabling the models to overlap with the observations. In addition, the upper bound of the initial mass function (IMF) is increased to allow sampling from the high mass (M $=120 ~\mathrm{M}_{\odot}$) model of LC18. These updated parameters allow the OMEGA$+$ code to reach a metallicity of [Fe/H]$\sim~0.6$ dex, ensuring our GCE models encompass the full metallicity range observed in the young thick disc sample. The metallicty adjustment was empirically optimised by varying the \textit{mgal} and \textit{sfe} parameters (see Table \ref{tab:stellar_params}) from their default values. Specifically, the \textit{sfe} parameter was tested over the range \(3.0\times 10^{-8}\) to \(3.0\times 10^{-11} \), with the final adopted value of \(3.0\times 10^{-10}\) providing the best agreement with our science objectives.
%%%%%%%%%%%%%%%%%%%%%%%%%%%%%%%%%%%%%%%%%%%%%%%%%%
\subsubsection{Stellar Nucleosythesis Yields}\label{sec:stellar_yields}
For all GCE simulations, nucleosynthetic yields constitute an integral ingredient. The yields consist of materials expelled by each isotope from the different nucleosynthesis channels. The nucleosynthesis yields adopted in our GCE models represent the total mass of chemical species expelled by a star over its lifetime. These yields are computed following the methodology presented in \citet{Karakas2007}, using the equation:
\begin{center}
\begin{equation}
 M_i = \int_{0}^{\tau} [X(i) - X_0(i)]\frac{dM}{dt}\,dt, 
  \end{equation}
\label{eq:stellar_yield}
 \end{center}
where $M_i$ denotes the yield of a given species, $\frac{dM}{dt}$ is the stellar mass loss rate at time $t$, $X(i)$ and $X_0(i)$ are the current and total mass fractions of the species, respectively, and $\tau$ is the total lifetime of the stellar model. A positive yield indicates the net production of the species during the star's evolution, whereas a negative yield signifies destruction relative to the star's initial abundance.

Our study incorporates several yield datasets covering various stellar masses and metallicities. For low-- to intermediate--mass stars (AGB stars) with metallicities up to Solar (Z$_\odot = 0.02$, \citealt{Lodders2009}), we adopt yields from \citet[][hereafter K10]{Karakas2010}, covering masses between $1.0$ and $6.0~M_\odot$ and metallicities at Solar and below--Solar (i.e., Z $=0.0001, 0.008,  0.004 ~\text{and}~ 0.02$). Additionally, new high-metallicity ($Z$) AGB yields from \citet[][hereafter C22]{Cinquegrana2022} have been included, spanning metallicities in the range $0.04 \leq \text{Z} \leq 0.1$ and stellar masses from $1.0$ to $8.0~M_\odot$ in increments of $0.5~M_\odot$. Using the relation \(\mathrm{[Fe/H]} \approx \log_{10}\left(\frac{Z}{Z_{\odot}}\right)\), the metallicity range from C22 corresponds approximately to $0.3 \leq \mathrm{[Fe/H]} \leq 0.7$.

For massive stellar yields (from CC SNe), we use yield data from three different studies: \citet[][hereafter WW95]{Woosley1995}, \citet[][hereafter K06]{Kobayashi2006}, and \citet[][hereafter LC18]{Limongi2018}. Each CC SN model listed in Table \ref{tab:model} adopts a distinct explosion mechanism, significantly influencing GCE predictions. Specifically, the K06 model assumes a piston-driven explosion with a fixed explosion energy, typically around $ 1.2\times10^{51}$erg. The WW95 model employs the KEPLER hydrodynamic code, also using a piston-driven mechanism, but explores a wider range of explosion energies ($0.3-3.0\times10^{51}$ erg) without including mixing processes. Finally, the LC18 model utilises hydrodynamic simulations with a kinetic bomb mechanism, incorporating variable explosion energies, variable remnant mass cuts, and feedback effects.
The WW95 yields, which specifically cover stellar masses of $13$, $15$, $18$, $20$, $25$, $30$, and $40$ Solar masses, have been included because our preliminary results from Figure \ref{fig:figure2} suggest that these yields produce enhanced Na abundances at higher metallicities. We observe that the mass range for WW95 is capped at $40$ solar masses, in contrast to the IMF adjusted value of $130$ Solar masses. This discrepancy will likely lead to potential underrepresentation of Na nucleosynthesis contributions from CC SNe from very high-mass stars. 

Lastly, for Type Ia supernovae (SNe Ia), we use yields from \citep[][, P22]{Pakmor2022}, computed at Solar metallicity (Z $= 0.02$) based on their `two-explosion' model of two merging white dwarfs and \citep[][S13]{Seitenzahl2013} Chandrasekhar-mass explosion. Although the choice between the single and double explosion models has minimal impact on the overall GCE results (see Figure \ref{fig:appendix}), we consistently apply either the S13 Chandrasekhar-mass or the P22 sub-Chandrasekhar-mass explosion model in our simulations. Table \ref{tab:model} summarises the adjusted values used for the GCE prediction for each model label in this study, and Table \ref{tab:raw_value} lists the raw Na and iron yield for the different categories of SNe Ia and CC SNe used in this study.
%%%%%%%%%%%%%%%%%%%%%%%%%% 
 \begin{table}
    \centering
    \caption{Yields in \(M_\odot\) of Na-23 and Fe-56 isotopes for CC~SNe and the two Type Ia supernova explosion models used in this work: Chandrasekhar-mass (\(M\mathrm{Ch}\)) model from S13 and sub-Chandrasekhar-mass (\(\text{sub-}M\mathrm{Ch}\)) model from P22 (see Table \ref{tab:model}). \textbf{Note:} Solar metallicity yields are shown, 
    although yield values for additional metallicities are included from S13 in our GCE modelling (i.e., \(Z = 0.0002, ~ 0.002, ~ \text{and}~ 0.01\)). The integrated yield per solar mass for different massive star yields used in this work has been computed and normalised using the \citet{Salpeter1955} IMF ($\alpha = 2.35$, at Z $= 0.02$).}
    \label{tab:raw_value}
    % \resizebox{\columnwidth}{!}{%
%   \begin{tabular}{lll}
%     \hline
%     SN\,Ia type      & ${}^{23}\mathrm{Na}$       & ${}^{56}\mathrm{Fe}$      \\
%     \hline
%     $M_{\rm ch}$     & $3.74\times10^{-5}$        & $6.22\times10^{-1}$       \\
%     sub-$M_{\rm ch}$ & $1.99\times10^{-4}$        & $4.87\times10^{-1}$       \\
%     \hline
%   \end{tabular}%
%   }
% \resizebox{\columnwidth}{!}{%
% \begin{tabular}{llccc}
% \hline
% Source & Mass Range & IMF & $^{23}$Na & $^{56}$Fe \\
%        & ($M_\odot$) & Weighting & Yield ($M_\odot$) & Yield ($M_\odot$) \\
% \hline
% \multicolumn{5}{l}{\textit{Type Ia Supernovae}} \\
% Mch      & --      & --                      & $3.74 \times 10^{-5}$ & $6.22 \times 10^{-1}$ \\
% sub-Mch  & --      & --                      & $1.99 \times 10^{-4}$ & $4.87 \times 10^{-1}$ \\
% \multicolumn{5}{l}{\textit{Core-Collapse Supernovae (CCSNe)}} \\
% WW95     & 11--40  & $m^{-2.35}$             & $2.1 \times 10^{-5}$  & $8.3 \times 10^{-2}$  \\
% K06      & 13--30  & $m^{-2.35}$             & $1.8 \times 10^{-5}$  & $7.6 \times 10^{-2}$  \\
% LC18     & 15--120 & $m^{-2.35}$             & $3.4 \times 10^{-5}$  & $1.1 \times 10^{-1}$  \\
% \hline
% \end{tabular}
% }
%current table
\resizebox{\columnwidth}{!}{%
\begin{tabular}{lccc}
\hline
Model & Mass Range ($M_\odot$) & ${}^{23}\mathrm{Na}$ ($M_\odot$ per event) & ${}^{56}\mathrm{Fe}$ ($M_\odot$ per event) \\
\hline
WW995    & 11--40    & $8.454\times10^{-5}$ & $1.941\times10^{-3}$ \\
K06    & 13--40    & $1.049\times10^{-4}$ & $1.742\times10^{-3}$ \\
LC18   & 13--120   & $8.406\times10^{-5}$ & $1.603\times10^{-3}$\\
\hline
SN\,Ia ($M{\rm Ch}$)       & ---       & $3.74\times10^{-5}$  & $6.22\times10^{-1}$ \\
SN\,Ia (sub-$M{\rm Ch}$)   & ---       & $1.99\times10^{-4}$  & $4.87\times10^{-1}$ \\
\hline
\end{tabular}
}
\end{table}
%%%%%%%%%%%%%%%%%%%%%%%%%%

%%%
\section{Results}
In this Section, we compare the GALAH DR3 observational data for thick-disc Solar-type stars with our GCE model predictions obtained using OMEGA$+$ and adjusted stellar yields. In short, we find that none of our GCE models reproduce the upturn in [Na/Fe], with this ratio instead declining for [Fe/H] $>$~0.
% \end{figure*}
% %%%%%%%%%%%%%%%%%%%%%%%%%%%%%%%%%%%%%%%%%%%%%% Figure 4
\begin{figure*}  
  \centering
  % First panel (Figure 3)
  \begin{subfigure}[b]{\textwidth}
    \centering
    % Limit its height to 0.45 of the text height; adjust if you need more/less.
    \includegraphics[width=\textwidth,
                     height=0.45\textheight,
                     keepaspectratio,
                     trim=5 10 5 10,   % (left bottom right top) in bp; tweak to crop margins
                     clip
                    ]{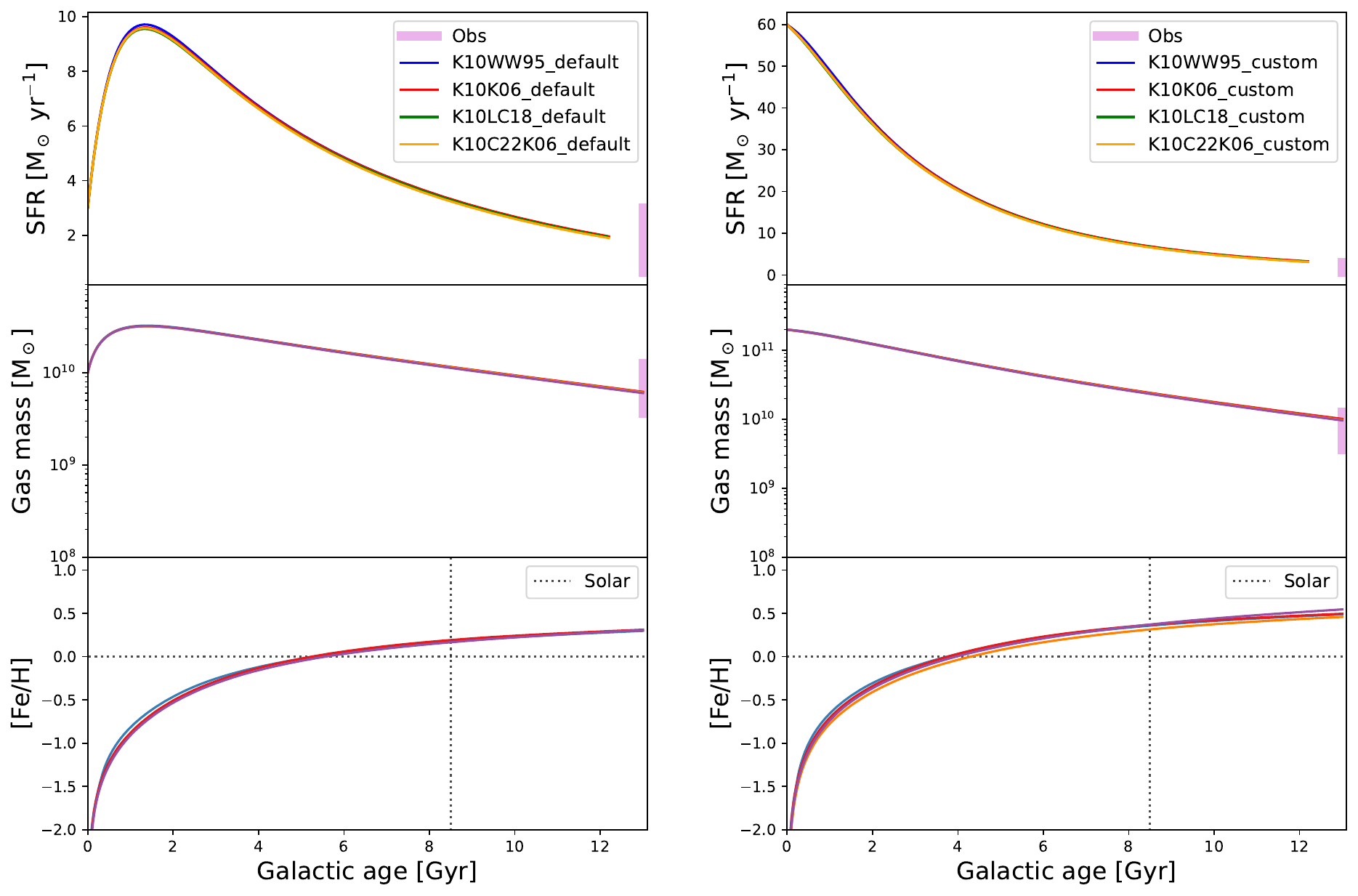}
    \caption{Calibration diagnostics for the Milky Way (MW) Galaxy model, using both default and adjusted input parameters (see Table \ref{tab:stellar_params}). Each panel presents model predictions for specific Galactic parameters, with vertical shaded bands (pink) representing observational constraints from \citet{Kubryk2015}. For calibration purposes only, the Chandrasekhar-mass Type Ia supernovae of \citet{Seitenzahl2013} were adopted for both default and adjusted models.}
    \label{fig:MW_calibration}
  \end{subfigure}
  \vspace{4pt}  
  % Second panel (Figure 4)
  \begin{subfigure}[b]{\textwidth}
    \centering
    \includegraphics[width=\textwidth,
                     height=0.45\textheight,
                     keepaspectratio,
                     trim=5 10 5 10,
                     clip
                    ]{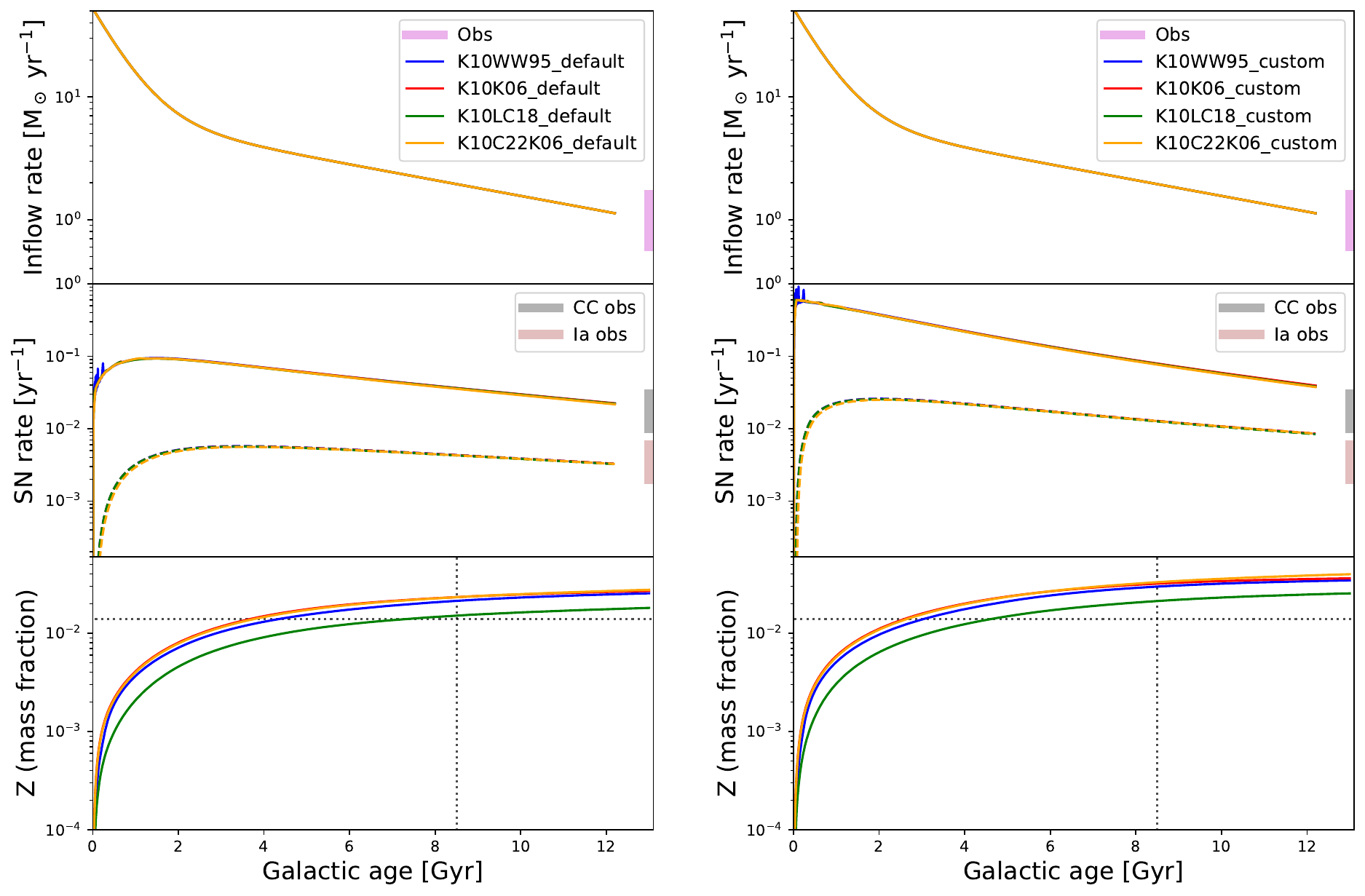}
    \caption{Comparison between default and adjusted calibration of the MW using parameters from Table \ref{tab:stellar_params}. Model labels follow those defined in Table \ref{tab:model}  across various diagnostic parameters.}
    \label{fig:customise_calibration}
  \end{subfigure}

  \caption{Bulk galactic properties of our model Galaxy based on adjusted model parameters.}
  \label{fig:fig4}
\end{figure*}

%%%%%%%%%%%%%%%%%%%%%%%%%%%%%%%%%%%%%%%%%%%%%%%%
\subsection{Milky Way model comparison}
\label{sec:MWMC}
Figure \ref{fig:fig4} (i.e., \ref{fig:MW_calibration} and \ref{fig:customise_calibration}) illustrates the Galactic properties used to validate our baseline Milky Way model, based on the values in Table \ref{tab:stellar_params}. The outcome as a result of the adjustment in SFE and \texttt{mgal} resulted in higher SFR as seen in Figure \ref{fig:MW_calibration} -- with the adjusted parameters having a higher SFR at earlier epochs ($\approx 60 \mathrm{~M_{\odot}~yr^{-1}}$) compared to the default parameters ($\approx 9 \mathrm{~M_{\odot}~yr^{-1}}$). The increase in star formation creates greater metal production, which contributes to greater [Fe/H] values over time. The substantial initial gas mass promotes vigorous early star formation, accelerating chemical enrichment \citep{Krumholz2012}. The elevated SFRs from the adjusted initial gas conditions align with the equilibrium frameworks of \citet{Krumholz2012}, showing that excess gas accretion can temporarily boost SFRs until equilibrium is reached \citep{Tinsley1978}.
 
The adjusted configuration shows higher [Fe/H] as a function of stellar age, reflecting a universal relationship between metallicity and the stellar--to--gas mass ratio. \citet{Zahid2014} demonstrated that higher initial gas masses accelerated metal production, directly raising [Fe/H] over time. This supports the concept of `gas processing equilibrium,' where metal yields are governed by the balance between gas accretion and star formation \citep{Tinsley1978, Koppen1999}. Consequently, it is expected that the adjusted parameters produce higher metallicity environments at `modern' galaxy ages.
%%%%%%%%%%%%%%%%%%%%%%%%%%%%%%%%%% Figure 5
\begin{figure*}
	\includegraphics[width=\textwidth]{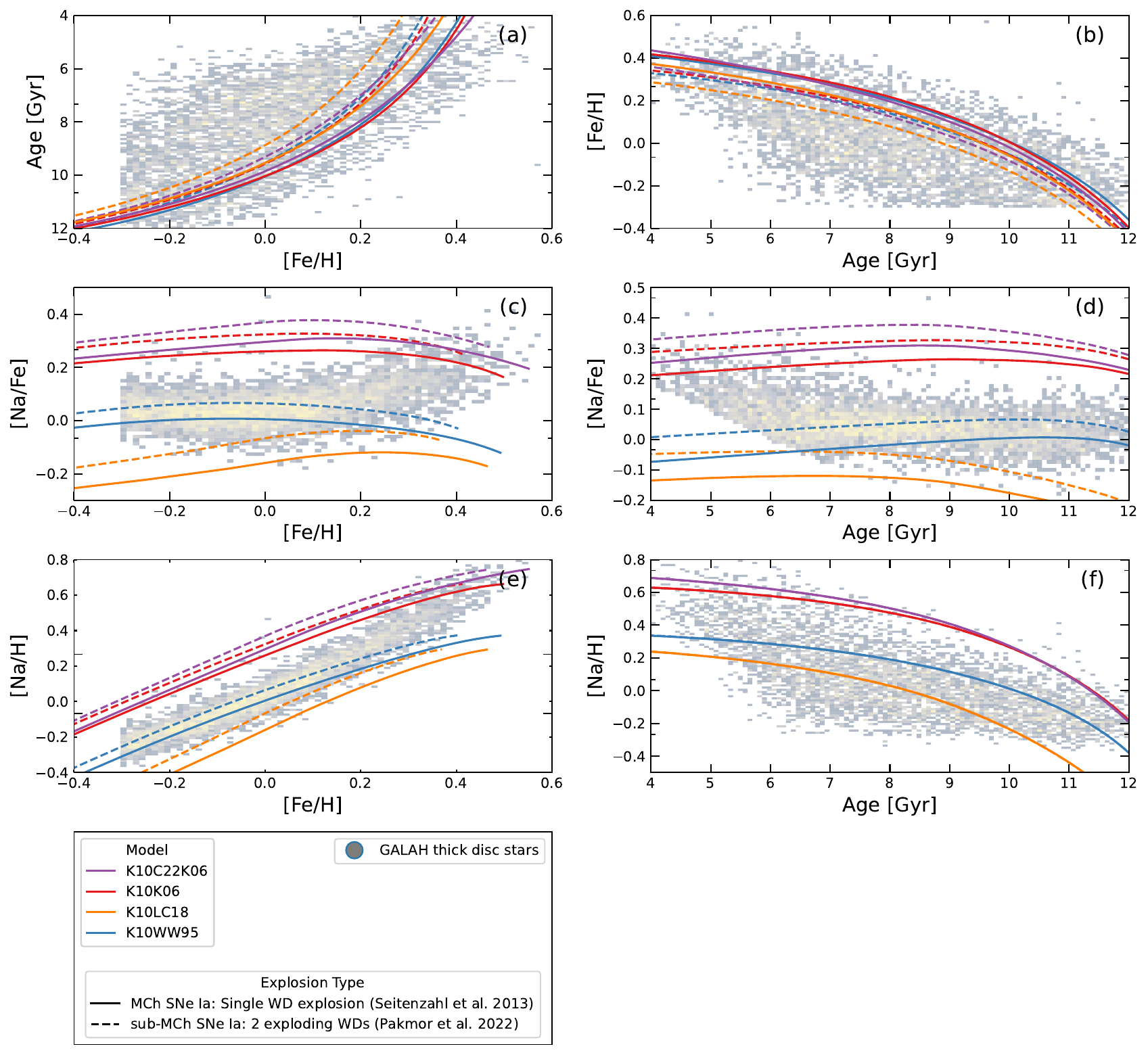}
     \caption{Panels (a)-(f) are the [Fe/H]-age, age-[Fe/H], [Fe/H]-[Na/Fe], age-[Na/Fe], [Fe/H]-[Na/H] and age-[Na/H] planes, produced by our adjusted GCE code parameters for this study. `Age' here refers to stellar ages. The grey bins are Solar-type young thick disc stars selected from the GALAH DR3 catalogue using Equation \ref{eq:Z_range}, with the yellower area having a higher concentration of stars. Solid lines represent GCE models in which all SN~Ia explosions are assumed to be from Chandrasekhar mass WDs, while for the dashed lines, sub-Chandrasekhar mass explosions from WD mergers are assumed. We defined the model label in Table \ref{tab:model}. SN Ia contribution is based on the two-exploding white dwarf from \citep{Pakmor2022} and delayed detonation for Chandrasekhar-mass white dwarfs, as described in \citep{Seitenzahl2013}.}
  \label{fig:figure5}
\end{figure*}
%%%%%%%%%%%%%%%%%%%%%%%%%%%%%%%%%%%%%%%%%%%%%%%%%%%%%%%%%%%%%%%%%%%%%%%%%%%%%%%%%
In the upper panels of Figure \ref{fig:customise_calibration}, the inflow rates for both default and adjusted models remained unchanged. The inflow rates span roughly \(1-10~\mathrm{M}_{\odot}\)/yr, reflecting standard assumptions regarding gas accretion from the circumgalactic medium for Milky Way-like galaxies \citep{Sanchez2014}. For SNe rates as a function of Galactic age, the CC SNe rates display pronounced initial peaks at early Galactic ages (around $1 - 4$ Gyr) in both default and adjusted models, with the adjusted model displaying enhanced rates, a consequence of the larger mass and SFR. In addition, all models (default and adjusted) agree reasonably well with the observational data at $\sim13$ Gyr. Finally, the metallicity evolution (i.e., mass fraction, $Z$) as a function of age for all models demonstrated the expected monotonic increase in $Z$. The default parameters achieve modest metallicity enrichment, rising gradually. 
%%%%%%%%%%%%%%%%%%%%%%%%%%%%%%%%%%%%%%%%%%%%%%%%%%%%%%%%%%%%%
\subsection{Abundance, age, metallicity plane}
Here, we describe the different results based on Figure \ref{fig:figure5}. In all panels, the data are those described in Section \ref{sec:data}. 
\begin{itemize}
\item[$\bullet$] \textbf{[Fe/H] vs Age (panel a):}
All models show a general trend of increasing [Fe/H] with decreasing stellar age. In particular, the K10C22K06 and K10K06 models closely follow each other, reaching higher [Fe/H] values at younger ages compared to the K10LC18 and K10WW95 models. This hints at CC SNe being the key ingredient in these simulations. The observational data show a spread in [Fe/H] across all ages, with a slight trend of higher [Fe/H] at younger ages. The observed trend (spread) in the observational data is corroborated with the age metallicity relation (AMR) previously reported in other works \citep[e.g.,][]{Sellwood2002}. 

\item[$\bullet$] \textbf{Age vs. [Fe/H] (panel b):}
This panel presents the same information as panel (a) but with axes swapped. The data indicate a large scatter in [Fe/H] for stars of similar age.

\item[$\bullet$] \textbf{[Fe/H] vs [Na/Fe] (panel c):}
All models exhibit an initial increase in [Na/Fe] as [Fe/H] increases, followed by a general decline beyond solar metallicity, with the decrease occurring at different [Fe/H] values for each model. The K10WW95 model best matches the [Na/Fe] and [Fe/H] parameter space, but it does not display the observed enrichment trend at [Fe/H] $>0$. 

\item[$\bullet$] \textbf{Age vs [Na/Fe] (panel d):} %%%% got here 17th June, 2025.
Depending on the model, there is a general trend of decreasing [Na/Fe] with stellar ages older than around $9$ Gyr or so. The K10C22K06 model predicts higher [Na/Fe] values for younger stars compared to other models. The observational data show a flat trend in older stars until around $6.5$ Gyr, after which there is a slight increase in [Na/Fe] in younger stars. The increasing [Na/Fe] with decreasing stellar age is less obvious in these data than in the [Na/Fe] vs. metallicity parameter space (panel c). 
%data show a slight trend of increasing [Na/Fe] with age, with significant scatter
\item[$\bullet$] \textbf{[Fe/H] vs [Na/H] (panel e):}
Both models and observational data display a continued rise in the [Na/H] vs. [Fe/H] plane. The two models with CC SNe from K06 have a higher [Na/H] than the other two models with different CC SNe yields. For instance, the model with CC SNe from LC18 lies below the data, while WW95 more closely matches the data. 

\item[$\bullet$] \textbf{Age vs [Na/H] (panel f):}
Models predict decreasing [Na/H] with increasing stellar age. The K10C22K06 and K10K06 models show higher [Na/H] values for younger stars compared to K10LC18 and K10WW95. Also, observational data show a weak trend of decreasing [Na/H] with age, with considerable scatter.
\end{itemize}
We note that the contribution of sub-Mch SNe Ia from \cite{Pakmor2022} resulted in higher [Na/Fe] abundances across all models for the various age, metallicity and [Na/Fe] planes explored as shown in Figure \ref{fig:figure5}. In this sub-Chandrasekhar mass model, Na yields from SNe Ia are $5$ times higher than in the Chandrasekhar mass model of \citet{Seitenzahl2013}, while iron production is $\sim 22$ per cent lower. 
%%%%%%%%%%%%%%%%%%%%%%%%%%%%%%%%%%%%%%%%%%%%%%%%%%%%%%%%%%%%%%%%%%%%%%%%%%%
\section{Discussion}
The adjustment to the IMF in Table \ref{tab:stellar_params} affects how much Na is produced in the Galaxy, mainly because of the different amounts produced by CC SNe and SNe Ia. Specifically, CC SNe contribute on average $\sim 10^{-5} - 10^{-4} M_\odot$ per event, while some SN Ia make on average $2\times10^{-4} M_\odot$ Na per event. From Table \ref{tab:raw_value}, CC SN yields are typically lower, ranging from $\sim 0.8$ to $1\times 10 ^{-4}~M_\odot$ per event. On the other hand, SNe Ia exhibit more varied yields, with MCh explosions releasing about $3.74\times 10^{-5}$ $M_\odot$ per event and sub-MCh explosions producing significantly more, around $1.99\times 10^{-4}$ $M_\odot$ per event. Although individual sub-MCh SNe Ia can release comparable or even greater amounts of Na than single CC SNe (see GCE line predictions in Figure \ref{fig:figure5}) per event, their overall Galactic contribution remains comparatively small due to their lower occurrence rates relative to CC SNe \citep{Li2011}. This frequency disparity ensures that, cumulatively, CC SNe dominate Na enrichment in the Galaxy, despite their lower per-event yields. However, because SNe Ia are primary contributors of iron (Fe), their metallicity-dependent yields, currently modelled predominantly at Solar metallicty, could disproportionately influence [Na/Fe] ratios at $\mathrm{[Fe/H]} > 0$.

Even though our current models can show the overall pattern in [Na/Fe] for values between about $-0.4$ and $0$, none of the models can explain the increase in [Na/Fe] that we see in the data (Figure \ref{fig:figure5}c). Although both K10K06 and K10C22K06 models share the same CC SNe yield tables, the K10C22K06 model includes additional yields from AGB stars with higher metallicity. Hence, as expected, both models behave similarly, with the K10C22K06 scaling higher [Na/Fe] as a function of [Fe/H] compared to the other models. Both of these models predict significantly higher [Na/Fe] across the whole metallicity range explored, while the K10LC18 model predicts sub-Solar [Na/Fe] values throughout. We note, however, that the K10LC10 model shows a promising upturn in [Na/Fe], though its value begins to flatten out and then decrease at around [Fe/H] ${\sim} 0.2$ dex.

The K10WW95 model matches the observed trend of [Na/Fe] correctly in the lower metallicity range, but it shows a decrease in [Na/Fe] at higher metallicities compared to what we see in the data. A previous study by \citet{Eitner2020} examining manganese abundances ([Mn/Fe] vs. [Fe/H]; see their fig. 2) highlighted differences in the explosion mechanisms of core-collapse supernovae as a likely explanation for the discrepancies between their NLTE data (i.e., $42$ stars including red giants and FG dwarf stars taken from different studies) and model predictions. Similarly, we attribute differences in our GCE model behaviour to the different CC explosion mechanisms used in \citet{Woosley1995, Kobayashi2006} and \citet{Limongi2018}.

Prior work using massive star yields by \citet{Kobayashi2011, Kobayashi2020a} indicated a plateau around Solar metallicity with a slight bump increase at sub-Solar metallicities. In a similar study of G-dwarf stars, \cite{Matteucci2014} used massive star yields from \citet{Woosley1995} but also looked at low- and intermediate-mass stellar yields from \citet{Van1997} for values between \(-3.5\) and \(0.2\) in \(\mathrm{[Fe/H]}\). This resulted in a GCE prediction that appeared rather flat at Solar metallicity (see their fig. $2.22$). Comparing their result to our findings, our GCE prediction flattens briefly at Solar metallicities and then decreases. This difference is probably due to the different amounts of elements produced, the ways stars explode, how AGB stars are considered, and the different assumptions made in the GCE codes used in each study. In addition, we note that:

\begin{itemize}
\item[$\bullet$] Our observational sample represents only a portion of the Galaxy’s stellar disc population and does not fully capture the diversity of chemical evolution across the entire Galaxy. Note that our GCE modelling uses OMEGA$+$, which is calibrated to the Milky Way Galaxy as a whole.

\item[$\bullet$] The discrepancies also stem from the assumptions inherent in the GCE models, particularly in the treatment of nucleosynthesis yields and stellar feedback processes at high metallicities. For instance, we increased the initial mass of gas in the simulation from the default values. While this approach enables us to extend the model to higher [Fe/H] values, it somewhat diverges from a truly realistic representation of our Galaxy (see Figure \ref{fig:MW_calibration}).

\item[$\bullet$] The failure of our models to reproduce the observed Na enhancement relative to iron at super-Solar metallicities implies a potential lack of iron enrichment in these stars. In other words, the younger old thick disc Solar-like stars may have experienced a chemical evolution in which the progenitor `pre-stellar' cloud was under-abundant in pollutants expelled by Type Ia supernovae (e.g., iron). 
\end{itemize}

Among the models examined, the K10WW95 model most closely passes through the observational data until Solar values and then declines from around [Fe/H] $\sim 0.1$ dex instead of continuing the observed upturn in the data. In contrast, the K10K06 and K10C22K06 models tend to predict higher [Na/Fe] values at all metallicities sampled by the data. A key factor in the [Na/Fe] enrichment is the contribution from AGB stars, which produce Na through the Ne-Na cycle during HBB at low metallicity and second dredge-up at higher metallicities \citep{Mowlavi1999, Ventura2013, Karakas2014}. The efficiency of this process increases at higher metallicities due to enhanced neon abundances \citep{Karakas2014, Cristallo2015}. We note that the K10C22K06 model displays the most prominent increase in [Na/Fe] as a function of [Fe/H], though with our current GCE set-up, the overall [Na/Fe] values are well below those of the observations (Figure \ref{fig:figure5}, panel c). However, one could envisage a scenario in which SNe~Ia played a lesser role in polluting the interstellar gas that gave rise to forming stars in our sample. In such a case, iron would be under-abundant, which would naturally lead to an increase in the absolute abundance of [Na/Fe], plausibly bringing the models into better agreement with observations.

Notably, none of our tested models that combine massive and AGB stars and either Chandrasekhar- or sub-Chandrasekhar-mass SNe Ia successfully replicate the observed [Na/Fe] upturn at super Solar metallicities (see Figure \ref{fig:figure5}c). As a test, we tried two sets of SN~Ia yields from WD mergers: either the two-explosion model as already discussed, in addition to the single-explosion model of 
\citet{Pakmor2022} in which only the primary WD explodes. The models show no observable differences in [Na/Fe] predictions (Figure \ref{a:appendix}\ref{fig:appendix}). Collectively, these results suggest that the nature of the SN Ia progenitor does not explain the discrepancy between the models and the observations in metal-rich disc stars in this work. The physical mechanism behind this suppression remains unclear, though potential drivers include delayed chemical mixing in the thick disc or environmental factors altering (suppressing) SNe Ia progenitor evolution. While speculative, this last line of inquiry aligns most closely with observational trends and requires prioritising these in future GCE studies. 
%%%%%%%%%%%%%%%%%%%%%%%%%%%%%%%%%%%%

\section{Conclusions}
In this paper, we have examined Na enrichment in Solar-type Galactic young thick disc stars at super Solar metallicities, utilising observational data from the GALAH DR3 survey and testing GCE models that incorporate nucleosynthetic yields from massive stars, AGB stars (\(1.0 \leq M_{\odot} \leq 8.0\)) and two explosion mechanisms of SNe~Ia (Chandrasekhar-- and sub-Chandrasekhar mass explosions) using the OMEGA$+$ GCE code. At the time of writing, nucleosynthetic yields for sub-Chandrasekhar mass SNe Ia from WD mergers were only available for Solar metallicity and one set of WD merger masses. We acknowledge that future studies should aim to incorporate SN~Ia yields at a variety of metallicities (and masses in the case of sub-Chandrasekhar-mass scenarios) once they become more readily available. 

Critically, our analysis demonstrates that all four models incorporated in this study fail to replicate the [Na/Fe] enrichment at super-Solar metallicities. These findings suggest that our understanding of Na enrichment in metal-rich, young thick-disc, Solar-type stars remains insufficient, potentially reflecting gaps in the assumptions underpinning state-of-the-art nucleosynthesis prescriptions. The inability of these models to reproduce the observed trend -- despite their calibration with modern stellar physics -- implies that key processes, such as metallicity-dependent mass-loss rates in AGB stars, proton-capture reactions efficiencies in CC SNe, or the role of binary interactions.

While the K10WW95 model using WW95 massive stars initially appears promising -- in terms of matching the [Na/Fe] level up to Solar metallicity (see Figure \ref{fig:figure2}), modelling using the OMEGA$+$ GCE code reveals that the upturn trend deviates significantly at super Solar regimes (see panel (c) of Figure \ref{fig:figure5}). These findings indicate that attributing the super Solar [Na/Fe] upturn primarily to metallicity-dependent CC SNe yields \citep[e.g.][]{Arnould1978} reveals a challenging puzzle that remains unsolved. In addition, although SNe Ia produce more $^{56}$Fe compared to CC SNe, the result of the GCE prediction still fails to correctly replicate the Na enrichment in the Solar-type sample used in this study. The enrichment of Na at super-Solar metallicity caused by the secondary dredge-up effect in AGB star models was unable to account for the observed increase in [Na/Fe] in the data, despite including high-Z AGB yields in the K10C22K06 model. As a result, future research should focus on enhancing nucleosynthesis models over various metallicities from all sources to better align theoretical predictions with the observed [Na/Fe] trends in the Galactic disc and other stellar populations. In this sense, some discrepancies between GCE models and observational data for a sub-sample of nearby Galactic disc stars might even be expected and can provide valuable insights into the complexities of stellar evolution and chemical enrichment. Specifically, metallicity-dependent SN Ia yields -- including sub-Chadrasekhar mergers across the mass and metallicity range should be explored, although we acknowledge that this can be a computationally intensive exercise. However, these systems exhibit isotopic sensitivities, e.g., $\alpha$-elements and neutron-rich isotopes \citet{Travaglio2005}, but can be explored further. 

\section*{Acknowledgements}
The authors thank the anonymous referee for thoughtful comments that helped to improve this paper. This work was supported by the Australian Research Council Centre of Excellence for All-Sky Astrophysics in 3 Dimensions (ASTRO 3D) through project number CE170100013. Part of this research was funded through AJR's Australian Research Council Future Fellowship award number FT170100243. EKO, AJR and IRS also acknowledge the support of the Australia-Germany Joint Research Cooperation Scheme (Funding Application ID: 57654415), which facilitated our collaboration with Researchers from the Max Planck Institute for Astronomy in Heidelberg, Germany. 

\section*{Facilities}

\textbf{AAT with 2df-HERMES at Siding Spring Observatory:}
The GALAH Survey is based on data acquired through the Australian Astronomical Observatory, under programs: A/2013B/13 (The GALAH pilot survey); A/2014A/25, A/2015A/19, A2017A/18 (The GALAH survey phase 1), A2018 A/18 (Open clusters with HERMES), A2019A/1 (Hierarchical star formation in Ori OB1), A2019A/15 (The GALAH survey phase 2), A/2015B/19, A/2016A/22, A/2016B/10, A/2017B/16, A/2018B/15 (The HERMES-TESS program), and A/2015A/3, A/2015B/1, A/2015B/19, A/2016A/22, A/2016B/12, A/2017A/14, (The HERMES K2-follow-up program). This paper includes data provided by AAO Data Central (datacentral.aao.gov.au).
\textbf{\textit{Gaia}: } This work has made use of data from the European Space Agency (ESA) mission \textit{Gaia} (\url{http://www.cosmos.esa.int/gaia}), processed by the \textit{Gaia} Data Processing and Analysis Consortium (DPAC, \url{http://www.cosmos.esa.int/web/gaia/dpac/consortium}). Funding for the DPAC has been provided by national institutions, particularly the institutions participating in the \textit{Gaia} Multilateral Agreement. 
\textbf{Other facilities:} This publication makes use of data products from the Two Micron All Sky Survey \citep{Skrutskie2006} and the CDS VizieR catalogue access tool \citep{Vizier2000}.

\section*{Software}

The research for this publication was coded in \textsc{python} (version 3.7.4) and included its packages
\textsc{astropy} \citep[v. 3.2.2;][]{Robitaille2013,PriceWhelan2018},
\textsc{corner} \citep[v. 2.0.1;][]{corner},
%\textsc{galpy} \citep[version 1.6.0;][]{Bovy2015},
\textsc{IPython} \citep[v. 7.8.0;][]{ipython},
\textsc{matplotlib} \citep[v. 3.1.3;][]{matplotlib},
\textsc{NumPy} \citep[v. 1.17.2;][]{numpy},
\textsc{scipy} \citep[version 1.3.1;][]{scipy},
We further made use of \textsc{topcat} \citep[version 4.7;][]{Taylor2005};

%%%%%%%%%%%%%%%%%%%%%%%%%%%%%%%%%%%%%%%%%%%%%%%%%%
%%%%%%%%%%%%%%%%%%%%%%%%%%%%%%%%%%%%%%%%%%%%%%%%%%
\section*{Data Availability}
The data used for this study is published by \citet{Buder2021} and can be accessed publicly via \url{https://docs.datacentral.org.au/galah/dr3/overview/}
%%%%%%%%%%%%%%%%%%%% REFERENCES %%%%%%%%%%%%%%%%%%

% The best way to enter references is to use BibTeX:

\bibliographystyle{mnras}
\bibliography{references} 

%%%%%%%%%%%%%%%%%%%%%%%%%%%%%%%%%%%%%%%%%%%%%%%%%%

%%%%%%%%%%%%%%%%% APPENDICES %%%%%%%%%%%%%%%%%%%%%

\appendix \label{a:appendix}

\section{Exploring the effect of sub-Chandrasekhar SN Ia}
\begin{figure*}
	\includegraphics[width=\textwidth]{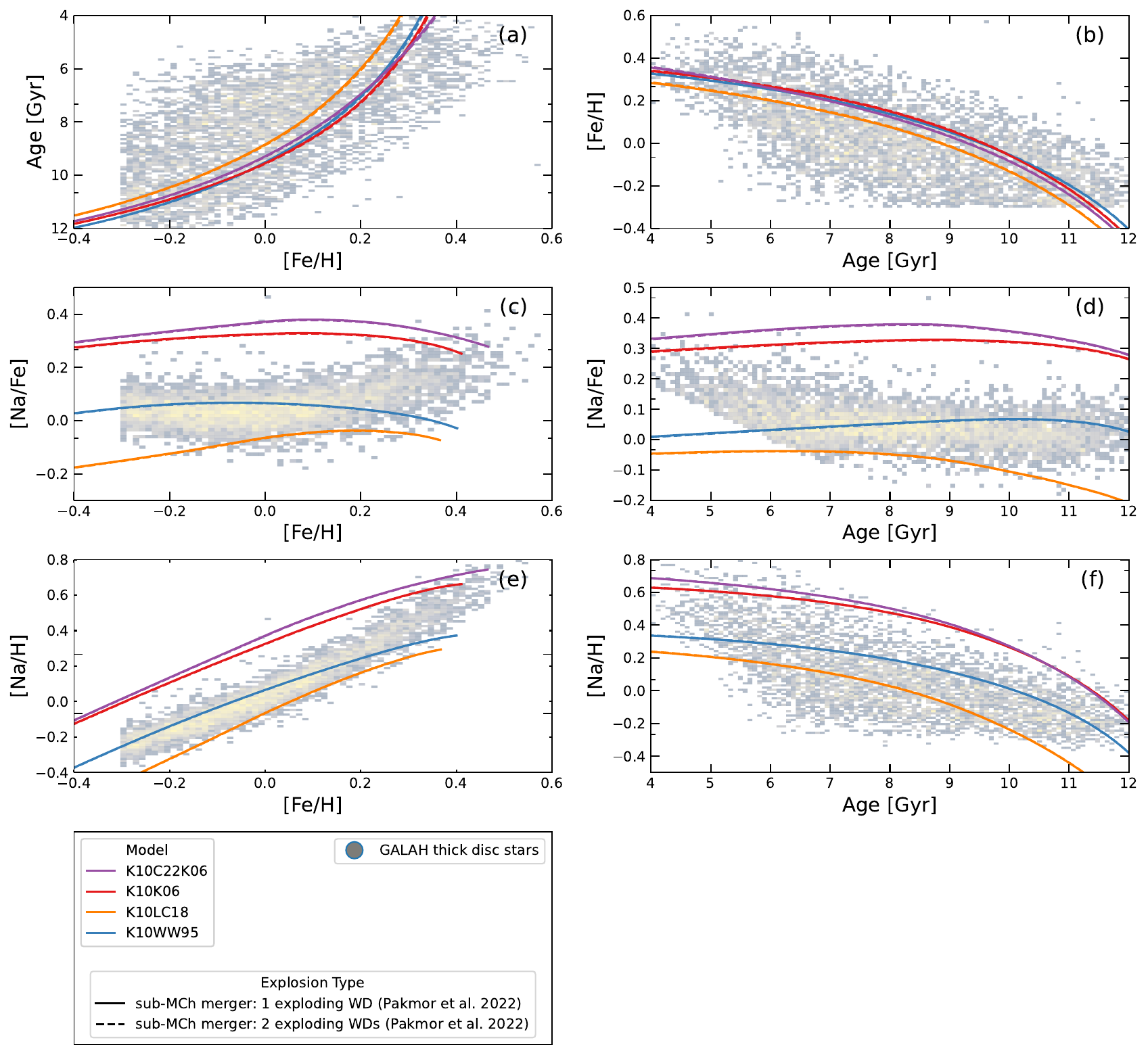}
     \caption{Na abundance relation for different parameters [Na/Fe], Age, [Fe/H] planes. We used the \citet{Pakmor2022} SNe Ia sub-Chandrasekhar ($1$ and $2$ explosion). For the GALAH data, yellower regions reflect a higher number density of stars.}
     \label{fig:appendix}
\end{figure*}
%%%%%%%%%%%%%%%%%%%%%%%%%%%%%%%%%%%%%%%%%%%%%%%%%%

% Don't change these lines
\bsp	% typesetting comment
\label{lastpage}
\end{document}